# Coarse Bifurcation Diagrams via Microscopic Simulators: A State-Feedback Control-Based Approach

Constantinos I. Siettos, Dimitrios Maroudas and Ioannis G. Kevrekidis

*Abstract*— The arc-length continuation framework is used for the design of state feedback control laws that enable a microscopic simulator trace its own open-loop coarse bifurcation diagram. The steering of the system along solution branches is achieved through the manipulation of the bifurcation parameter, which becomes our actuator. The design approach is based on the assumption that the eigenvalues of the linearized system can be decomposed into two well separated clusters: one containing eigenvalues with large negative real parts and one containing (possibly unstable) eigenvalues close to the origin.

## I. INTRODUCTION

A frequent feature of the solution branches of nonlinear systems depending on parameters is the occurrence of bifurcation points (such as turning points) involving exchange of stability. Numerical bifurcation theory is well established and used for the computation of such bifurcation diagrams in the modeling of nonlinear systems ranging form ecology to materials science and engineering. Open loop unstable steady states are typically found as fixed points of augmented sets of algebraic equations, through contraction mappings (like Newton-Raphson). Alternatively, assuming system controllability, an arsenal of existing controller design techniques can be used to achieve system stabilization. The interface between bifurcation theory and feedback control is an active area of current research. Representative examples include the local feedback stabilization and control of static and Hopf bifurcations of nonlinear systems [1,2,7]; bifurcation control, based on normal forms, via state feedback with a single uncontrollable mode [9]; the incorporation of washout filters [12, 20] to stabilize equilibria in the presence of model structural uncertainties; and global stabilization of bifurcations, using Lyapunov tools, changing the type of bifurcation from subcritical to supercritical [11]. All these approaches presume the availability of macroscopic mathematical models in closed form. However, it is often the case that macroscopic balance equations are not available in closed form, and the best available model is a microscopic/stochastic simulator (molecular dynamics, MD, kinetic Monte Carlo, KMC, kinetic theory based Lattice-Boltzmann, LB, or Markov chain simulators). Under these circumstances, conventional computational algorithms cannot be used explicitly for bifurcation analysis and controller design. Our group has established over the last few years [4, 14, 15, 18] that "coarse timesteppers" can serve as a bridge between microscopic / stochastic system descriptions and macroscopic tasks such as coarse bifurcation analysis and control computations.

In this paper we address the development of a feedback control scheme, which, implemented –under appropriate conditions- as a shell around existing microscopic simulators enables them to automatically trace, exploiting an arc-length parametrization, their "coarse" bifurcation diagram. Steering of the system (microscopic timestepper) along solution branches is achieved through the manipulation of the bifurcation parameter, which becomes our actuator. We demonstrate the results of implementing the proposed methodology for tracing the coarse (macroscopic) bifurcation branch of a KMC model of a surface reaction, which exhibits two turning points. The approach is conceptually also applicable to physical experiments.

Manuscript received January 31, 2003. This work was supported in part through AFOSR (Dynamics and Control, Dr. B. King) and an NSF/ITR grant.
C. I. Siettos is with the Department of Chemical Engineering, Princeton University, Princeton, NJ, 08544, USA (e-mail: csiettos@Princeton.edu).
D. Maroudas, was with the Department of Chemical Engineering, University of California, Santa Barbara, Santa Barbara, California, 93106, USA. He is now with the Department of Chemical Engineering, University of Massachusetts Amherst, Amherst, MA 01003-3110, USA (e-mail: maroudas@ecs.umass.edu)
I. G. Kevrekidis is with the Department of Chemical Engineering, Department of Mathematics and Program in Applied and Computational Mathematics, Princeton University, Princeton, NJ, 08544, USA (corresponding author, phone: (609)2582818, e-mail: yannis@Princeton.edu).

## II. BRANCH TRACING AND THE FEEDBACK SCHEME

We briefly describe this feedback control approach for the case continuous time, autonomous nonlinear systems of the type

$$\dot{x} = f(\mathbf{x}, p), f: \mathrm{R}^m \times \mathrm{R} \to \mathrm{R}^m \qquad (1)$$

where $x \in \mathrm{R}^m$ denotes the state vector, accessible through measurement, $p \in \mathrm{R}$ is the bifurcation parameter and $f$ is a smooth function. Consider the design of feedback control laws that stabilize (1) locally around an equilibrium ($x^*, p^*$), which however is characterized by uncertainty.

To cope with equilibrium uncertainty, dynamic state feedback is used (in contrast to static state feedback which would most likely be applied if the system equilibrium was precisely known beforehand). An additional state variable $q$ defined by

$$\dot{q}(t) = g(x, p), g: R^m \times R \to R \qquad (2)$$

is incorporated. Motivated by arc-length continuation [10], we choose $g(x, p)$ to be the linearized pseudo arc-length condition. Assuming that two equilibrium points $(x_0, p_0)$ and $(x_1, p_1)$ are already known from following the steady state solution branch, the condition, with pseudo arc-length step $\Delta S$, reads:

$$g(x, p) \equiv \frac{(x_1 - x_0)^T}{\Delta S}(x - x_1) + \frac{(p_1 - p_0)}{\Delta S}(p - p_1) - \Delta S \qquad (3)$$

These equilibrium points can be computed using two ways when working with a model to the computer: the first way is what we would have done in an experiment: we set initial and operating conditions, and "run" (evolve) in time until nothing visibly changes. The second way refers to numerical analysis assisted tools (contraction mappings like Newton-Raphson or so-called Newton-Picard timestepper based algorithms using Krylov subspace iterations (e.g. the Recursive Projection Method of Shroff and Keller, [16]) which help to extract information from models "easier, faster, better" than simple simulation (see also [13,19]).

The linearization of (1) is performed around the second equilibrium point $(x_1, p_1)$:

$$\Delta \dot{x} = \left.\frac{\partial f}{\partial x}\right|_{x_1, p_1} \Delta x + \left.\frac{\partial f}{\partial p}\right|_{x_1, p_1} \Delta p \qquad (4)$$

where $A \equiv \left.\frac{\partial f}{\partial x}\right|_{x_1, p_1}$ is the Jacobian of the system, $B \equiv \left.\frac{\partial f}{\partial p}\right|_{x_1, p_1}$ is the control matrix, $\Delta x \equiv x - x_1$, $\Delta p \equiv p - p_1$ denote the deviation state and control variables.

For the local stabilization of (1)-(3), the choice here is to employ a state linear feedback controller of the form

$$\Delta p = [k_1 \ k_2] \begin{bmatrix} \Delta x \\ q \end{bmatrix}, k_1 \in R^m, k_2 \in R \qquad (5)$$

The gain matrix $[k_1 \ k_2]$ is calculated on the basis of placing the critical (slowest) eigenvalues at certain prescribed values in the left half of the complex plane so that local stabilization of (1)-(3) is achieved. The control law (5) then drives the system (1)-(3) to a steady state, where $\Delta \dot{x} = 0$, $\dot{q} = 0$ ($t \to \infty$) and thus both the steady state and the arc-length conditions are satisfied.

Note that in general, both the Jacobian and the control matrix of the augmented system at the new steady state $(x^*, p^*)$ will be slightly, for small values of arc-length step $\Delta S$, different from those at equilibrium point $(x_1, p_1)$. A guess for the Jacobian and the control matrix at the uncertain equilibrium point $(x^*, p^*)$ is obtained by taking an Euler step of length $\Delta S$ along the tangent at $(x_1, p_1)$.

A basic assumption for the validity of the above design procedure is that the linearized augmented model (3)-(4) should be controllable or at least stabilizable; indeed, for systems undergoing only simple limit points, the augmented linearized system (3)-(4) is controllable.

From the Popov-Belevitch-Hautus (PBH) test for Controllability [9], a linearized system of the form $\dot{x} = A_{n \times n} x + B p$, is controllable if and only if rank $[sI\text{-}A; B] = n$ for all s.

Clearly this condition will be met for all s that are not eigenvalues of $A$, because then $\text{Det}(sI - A) \neq 0$. From the linearized system (3)-(4) the PBH test reads:

$$\text{rank}\begin{pmatrix} sI - A & 0 & B \\ a & s & \beta \end{pmatrix} = n+1, \text{ where}$$

$$\frac{\partial g(x, p)}{\partial x} = \frac{(x_1 - x_0)^T}{\Delta S} \equiv a \qquad (6a)$$

$$\frac{\partial g(x, p)}{\partial p} = \frac{(p_1 - p_0)}{\Delta S} \equiv \beta \qquad (6b)$$

But $\text{rank}\begin{pmatrix} sI - A & 0 & B \\ a & s & \beta \end{pmatrix} = \text{rank}\begin{pmatrix} sI - A & B & 0 \\ a & \beta & s \end{pmatrix}$.

The matrix $\begin{pmatrix} s\mathbf{I}-\mathbf{A} & \mathbf{B} \\ \mathbf{a} & \beta \end{pmatrix}$ is nonsingular at simple limit points [10].

Indeed: rank $\begin{pmatrix} s\mathbf{I}-\mathbf{A} & \mathbf{B} \\ \mathbf{a} & \beta \end{pmatrix}$ = $n$+1 and as a consequence rank $\begin{pmatrix} s\mathbf{I}-\mathbf{A} & \mathbf{B} & 0 \\ \mathbf{a} & \beta & s \end{pmatrix}$ = $n$+1. □

The design of discrete time control systems is similar in principle, and the procedure we described above can be applied for the stabilization of the corresponding discrete-time fixed points (fixed points of the time-T map arising from integrating the continuous time ODEs). It is in the discrete time context that microscopic simulators are "coarsely" stabilized.

The main assumption behind the concept of designing coarse controllers for microscopic simulators is that, if coarse closed models exist (but are unavailable in closed form) then the higher moments of microscopically evolving distributions are quickly slaved to a few, say ρ, lower moments of these distributions (refer to [4,14,17] for a more detailed discussion). Coarse controller design is based on the information extracted, with the aid of coarse timestepper, at the system level from microscopic models, sidestepping the derivation of an explicit, closed form macroscopic description. The coarse timestepper provides a framework for estimating on demand ("just in time" [3]) quantities we need from the unavailable macroscopic system equations (the right-hand-sides of equations, the action of slow Jacobians, coarse derivatives with respect to parameters etc.). In short, what a controller design algorithm would obtain from a macroscopic model through function evaluations, is now estimated through short, appropriately initialized "bursts" of microscopic simulation.

Coarse (macroscopic) steady states can be obtained as fixed points, using T as the sampling time, of the mapping $\boldsymbol{\Phi}_T$ under which $\boldsymbol{x}(k+1) = \boldsymbol{\Phi}_T(\boldsymbol{x}(k), p)$, $\boldsymbol{\Phi}_T: \mathrm{R}^m \times \mathrm{R} \to \mathrm{R}^m$. Here the vector $\boldsymbol{x}$ denotes the coarse statistics (typically zeroth- or first-order moments of the microscopically evolving probability densities) that could deterministically describe in an efficient manner the long-term macroscopic behavior of the system under study.

## III. SIMULATION RESULTS

Our illustrative model is a KMC "stochastic simulation algorithm" [5, 6] Monte Carlo realization of a simplification (7) of the kinetics of NO oxidation reaction by $H_2$ on Pt and Rh surfaces:

$$\dot{\theta} = \alpha(1-\theta) - \gamma\theta - k_r(1-\theta)^2\theta \qquad (7)$$

where $\theta$ is the coverage of adsorbed NO, $\alpha$ is the rate constant for adsorption, $\gamma$ is the rate constant for desorption of NO, and $k_r$ is the reaction rate constant. Simulation results were obtained at: $\alpha$ = 1, $\gamma$ = 0.01; $k_r$ is the bifurcation parameter (and, in our scheme, the control variable). The deterministic version of the model exhibits two turning points (at $k_r \approx 3.96$ and $k_r \approx 26$). KMC simulations approximate the solution of the corresponding master equation, which describes the evolution of the probability (PDF) of finding the system in a certain configuration. For the stochastic simulations the values of the number of available sites (system size) say N and $N_{run}$ (number of realizations) were chosen as 800x800 and 100 respectively.

The coarse timestepper $\theta(k+1) = \boldsymbol{\Phi}(\theta(k), k_r(k))$ of the KMC model is used as a "black box". The coarse, Jacobian and control matrix are estimated by wrapping a computational superstructure like Newton's method as a shell around the coarse timestepper (see [14]). The coarse identified model (Jacobian and control matrix) is then used in the controller design procedure we discussed to iteratively trace the solution branch. The bifurcation diagram is given in Figure 1.

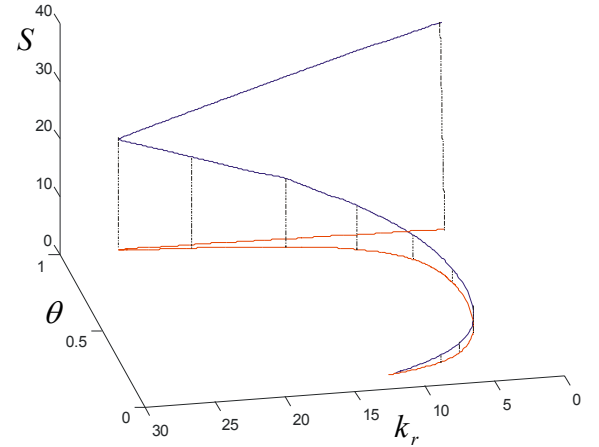

Fig. 1 Coarse bifurcation diagram of a KMC simulator based on eq.(7), obtained through pseudo-arclength (S) augmentation and feedback stabilization

The eigenvalue of the discrete time open loop system, as calculated from the estimated coarse Jacobian, is given in figure 2. The sampling time was set to T=0.5.

Figure 3 shows how, using two already found coarse steady states: $(\theta_0, k_{r0}) \approx (0.4657, 3.985)$ $(\theta_1, k_{r1}) \approx (0.4920, 3.9617)$, the procedure can drive the microscopic simulator to a (stabilized) unstable coarse steady state $(\theta^*, k_r^*) \approx (0.5157, 3.9617)$ beyond the turning point. The estimates of the Jacobian and control matrix of the unknown macroscopic equation computed by taking an Euler step at $(\theta_1, k_{r1})$, were 1.0405 and -0.063 respectively, while the values of *a* and β were: *a* =1.0544 and β = -0.9319. A value of ΔS=0.025 was chosen.

To perform the continuation, we placed the eigenvalues of the corresponding augmented linearized system, as obtained from eqs (3)-(4), to $\lambda_1$ =0.93, $\lambda_2$ = 0.91. The required gains were $k_1 \approx 0.0662$, $k_2 \approx -0.2196$.

The corresponding microscopic transients are shown in Fig. 4. Upon convergence the procedure is updated, a new stabilizing controller is designed, and the "next" point on the coarse branch is found.

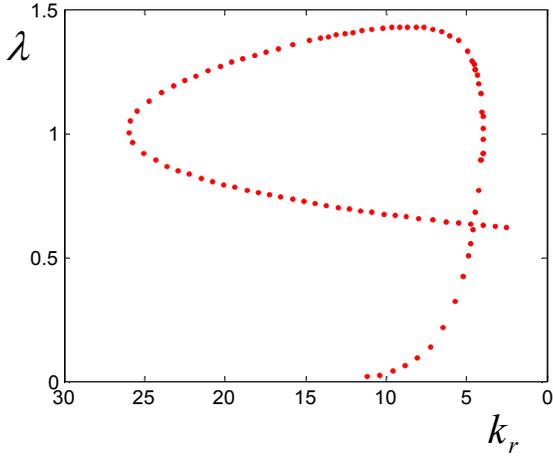

Fig. 2. Open loop eigenvalues of the steady states in Fig.1

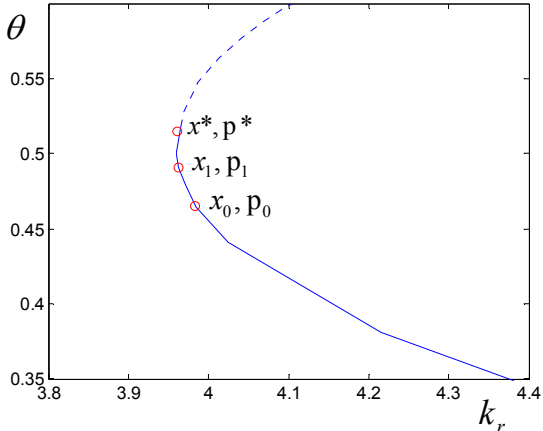

Fig. 3. Stabilization of a coarse stationary state $(x^*,p^*)$ beyond the turning point, with the help of two previously computed steady states $(x_0,p_0)$, $(x_1, p_1)$.

## IV. CONCLUSION

In summary, we have developed a feedback-control based scheme, which, under appropriate conditions, enables a microscopic/stochastic simulator to trace its "coarse" bifurcation diagram. This procedure might be helpful in the analysis of systems for which microscopic simulators are available, but no coarse macroscopic equations exist in closed form.

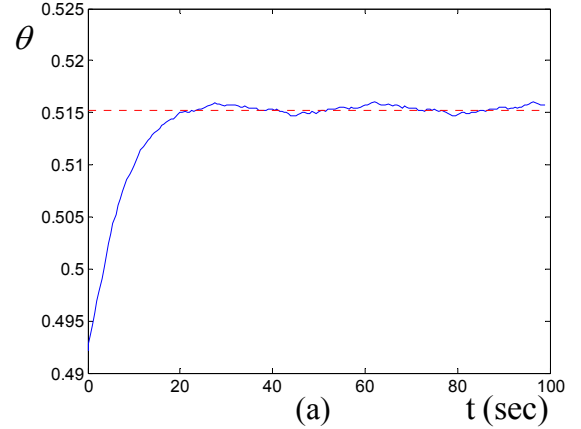

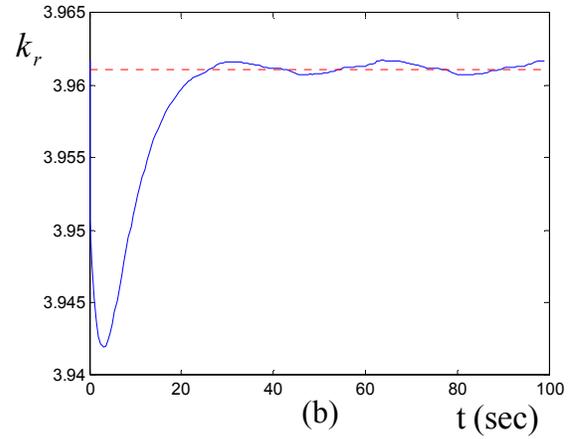

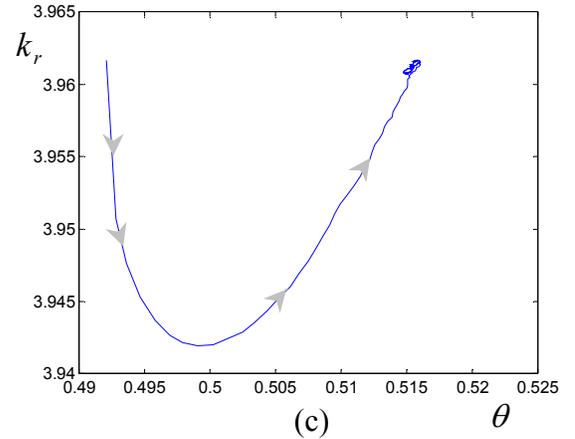

Fig. 4 Closed loop transients (a, b) and phase portrait (c) of the approach of the coarse state $\theta$ and the control action $k_r$ to a new (open-loop unstable) coarse steady state of the microscopic simulator past the turning point (see Fig.3).

ACKNOWLEDGMENT

We would like to acknowledge extensive discussions and collaboration with Prof. Alexei Makeev of Moscow State University.

…